\documentclass[aps,prl,twocolumn,groupedaddress,amsmath,amssymb]{revtex4-1}
\usepackage{graphicx}  
\usepackage{dcolumn}   
\usepackage{bm}        
\usepackage{verbatim}   

\begin{document}

\title{Reducing phase singularities in speckle interferometry by coherence tailoring}
\author{Klaus Mantel}
\affiliation{   
   Max Planck Institute for the Science of Light,\\
   Staudtstr 2, 91058 Erlangen, Germany\\
   }
\author{Vanusch Nercissian}
\affiliation{ 	 
   Institute of Optics, Information, and Photonics,\\
   Friedrich-Alexander-University Erlangen-N\"urnberg (FAU),\\
   Staudtstr. 7/B2, 91058 Erlangen, Germany\\
   }
   
\date{\today}
\begin{abstract}
Speckle interferometry is an established optical metrology tool for the characterization of rough objects. The raw phase, however, is impaired by the presence of phase singularities, making the unwrapping procedure ambiguous. In a Michelson setup, we tailor the spatial coherence of the light source, achieving a physical averaging of independent, mutually incoherent speckle fields. In the resulting raw phase, the systematic phase is preserved while the number of phase singularities is greatly reduced. Both interferometer arms are affected by the averaging. The reduction is sufficient to even allow the use of a standard unwrapping algorithm originally developed for smooth surfaces only.  
\end{abstract}

\maketitle

Speckle interferometry is a well established technique to characterize rough objects by optical means \cite{Mal2007}. Owing to the multiple beam interference generated by the rough surface, the systematic phase coming from the surface is drowned in a random phase caused by the surface roughness. In contrast to interferometry for smooth surfaces, an additional independent measurement is needed. When the measurement results are suitably combined, the resulting (combined) raw phase contains the desired information about the specimen.  This may be done with the help of a second wavelength \cite{Ferch1985}, so that the raw phase indicates the surface deviations of the specimen. Keeping a single wavelength, a second measurement in a different object state, e. g. after applying a load, provides information about the deformation of the specimen caused by the load \cite{Volk1980}. 

\begin{figure}[htbp]
\centering
\fbox{\includegraphics[width=0.90\linewidth]{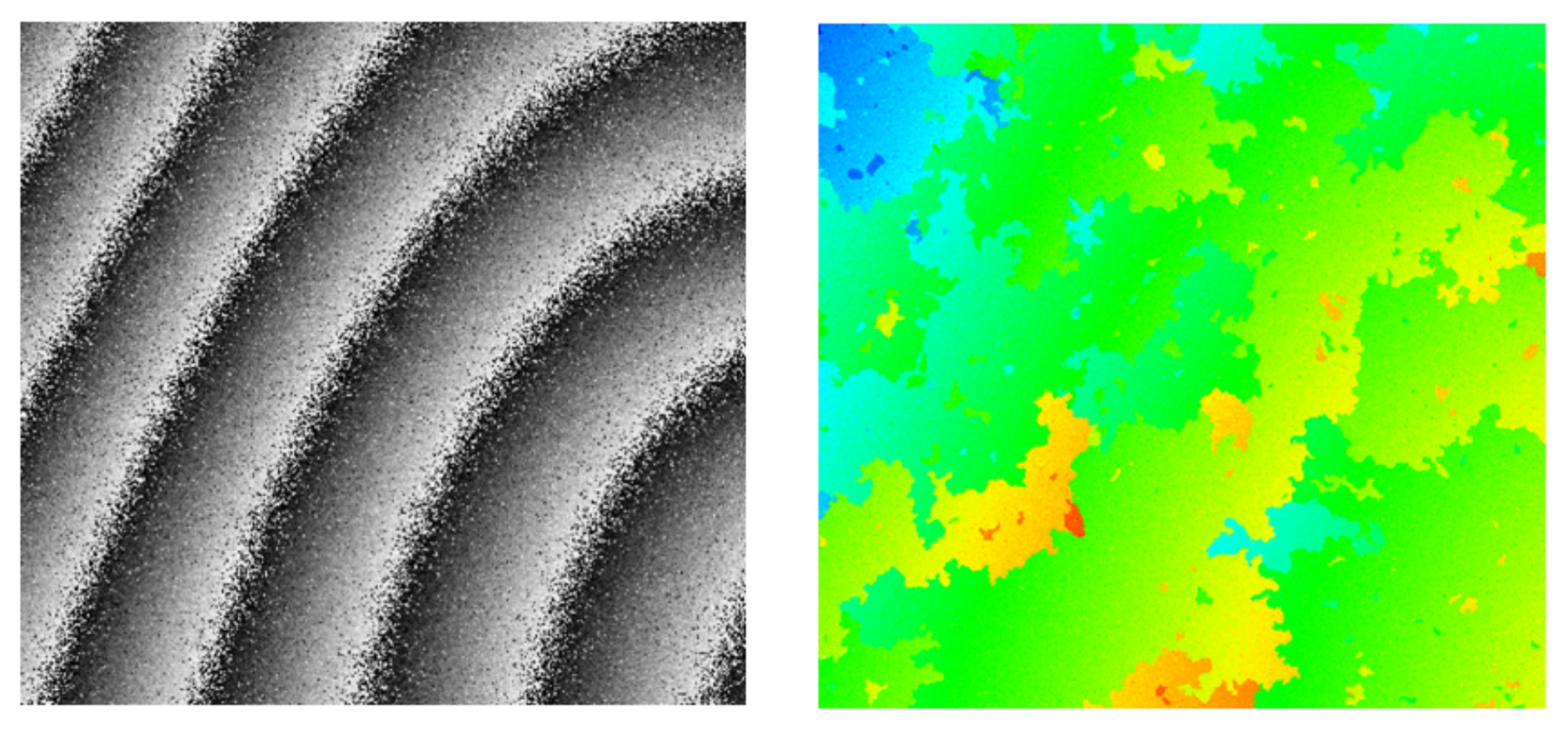}}
\caption{Results of a speckle deformation measurement. Left: Raw phase indicating a deformation of the specimen (values in the interval (-$\pi$,$\pi$]). Right: Unwrapped phase, obtained with a standard unwrapper for smooth surfaces. The high number of phase singularities in the raw phase make the unwrapping ambiguous and the unwrapping procedure fails.} 
\label{lExpPointSource}
\end{figure}

The raw phase is however impaired by the presence of phase singularities \cite{Nye1974}. Such phase singularities are an unavoidable consequence of the multiple beam nature of the interference from a rough surface. When trying to unwrap such a raw phase by a standard method \cite{Gig1998}, i. e. by an unwrapping algorithm originally developed for smooth surfaces \cite{Schr1998}, the result is extremely poor. Figure \ref{lExpPointSource} shows such a raw phase and the corresponding unwrapping result. No traces of the systematic phase coming from the deformation of the specimen are visible, instead the result is dominated by unwrapping errors.

A lot of work has been done to develop unwrapping algorithms that can overcome this problem \cite{Hunt2001}. Often, filtering and smoothing operations are performed, linear as well as nonlinear. On the other hand, it would be much preferable if the number of phase singularities in the raw phase were much lower in the first place. This reduction of the phase singularities should be achieved by a physical process, and not by software.

\noindent

\begin{figure}[htbp]
\centering
\fbox{\includegraphics[width=0.75\linewidth]{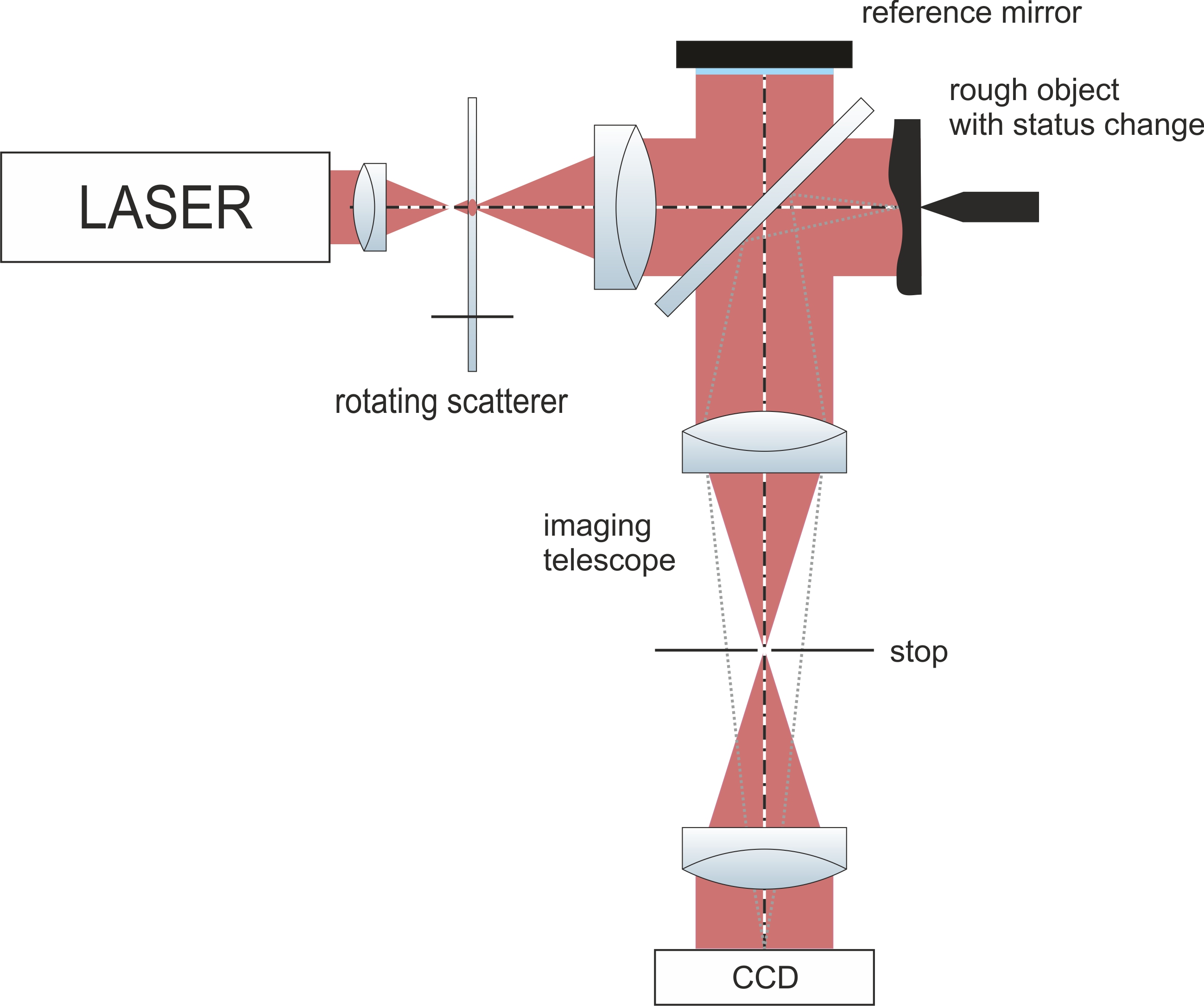}}
\caption{Experimental setup for a deformation measurement with reduced number of phase singularities. The single point source is replaced by an extended, incoherent source, generated by a defocused light spot on a rotating scatterer.} 
\label{lSketchDefDisc}
\end{figure}

Such a physical process is the incoherent averaging of several independent, mutually incoherent speckle fields \cite{Good1976},\cite{Crea1985}. Here, we tailor the spatial coherence of the light source to generate the averaging \cite{Mantel2016}. Since both interferometer arms are equally affected, the systematic phase is preserved while the phase singularities at the same time `average out'. Figure \ref{lSketchDefDisc} illustrates the procedure in case of an out of plane deformation measurement. A laser spot falls slightly defocused on a rotating scatterer. The amount of defocus determines the size of the spot on the scatterer. The points of this extended, effective light source can then be assumed to be effectively incoherent. The effective light source is then collimated and illuminates the interferometer. The phase is recovered via phase shifting using the reference mirror \cite{Bruning1974}.

\noindent

\begin{figure}[htbp]
\centering
\fbox{\includegraphics[width=0.90\linewidth]{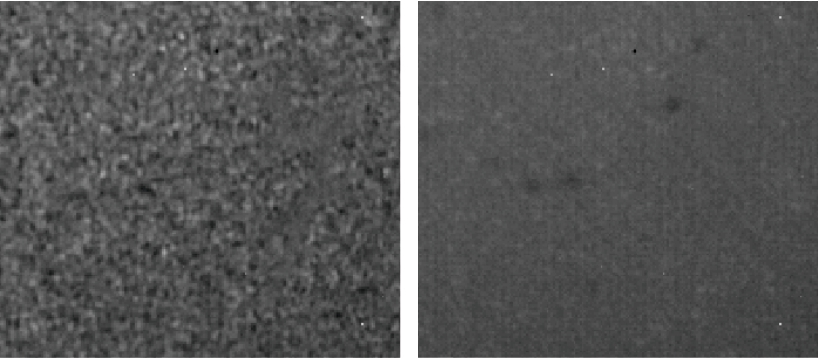}}
\caption{Speckle patterns for an extended, incoherent light source (experimental result). Left: interferometer arms balanced. Right: imbalanced interferometer arms, OPD$\approx$2cm.} 
\label{lExpSpecklePattDisc}
\end{figure}

From the interferometric testing of smooth surfaces, it is known that in case of spatially partially coherent illumination, the interference phenomenon is located around the common vertex of both the reference mirror and the specimen \cite{Schwid2006}. Correspondingly, here a high contrast speckle pattern can be observed when the reference mirror and the specimen are balanced, i. e. they are virtually in the same plane. This is shown in Fig. \ref{lExpSpecklePattDisc}.   

\noindent

\begin{figure}[htbp]
\centering
\fbox{\includegraphics[width=0.90\linewidth]{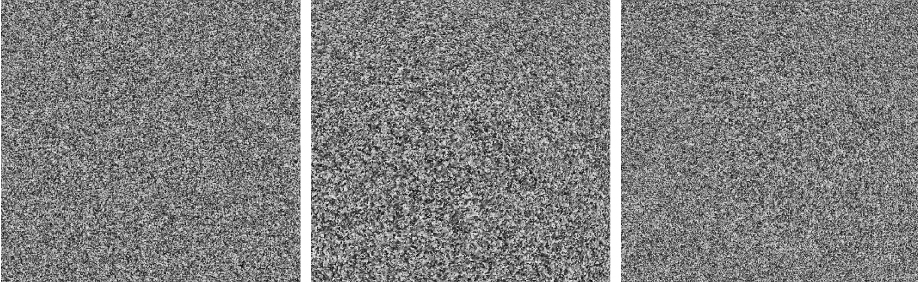}}
\caption{Speckle sizes (raw phase) for an extended, incoherent light source with varying diameter (experimental result). Left: Small source. Middle: extended light source. Right: Even larger extended light source.} 
\label{lExpSpeckleSizesDisc}
\end{figure}

Furthermore, it can be observed in Fig. \ref{lExpSpecklePattDisc} that the incoherent averaging has a direct effect on the speckle size, the speckle being larger for the extended light source compared to the point source. This already leads to a reduction of the number of phase singularities in the single measurements. The effect is, however, not linear, as Fig. \ref{lExpSpeckleSizesDisc} shows. For a certain light source diameter, the speckle size reached a maximum, only to be decreasing with further growing light source extension. It was at this maximum where the following measurements were taken.

\noindent

\begin{figure}[htbp]
\centering
\fbox{\includegraphics[width=0.90\linewidth]{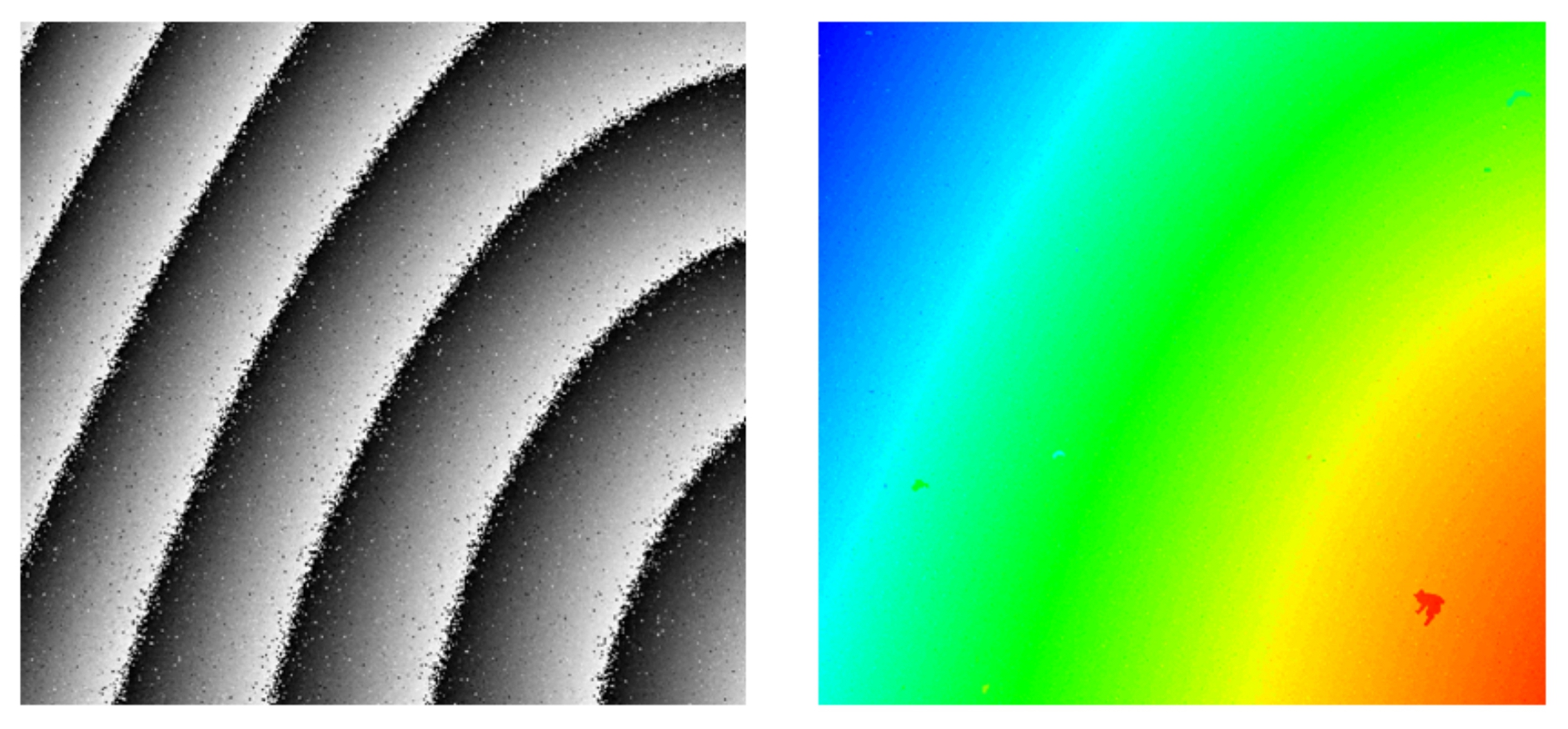}}
\caption{Results of a speckle deformation measurement, physical averaging applied (extended light source, experimental result). Left: Raw phase (values in the interval (-$\pi$,$\pi$]). Right: Unwrapped phase (pv 7.53$\lambda$), obtained with the same standard unwrapper for smooth surfaces. The number of phase singularities in the raw phase is drastically reduced, allowing for an almost error free unwrapping.} 
\label{lExpRedDisc}
\end{figure}

\begin{figure}[htbp]
\centering
\fbox{\includegraphics[width=0.95\linewidth]{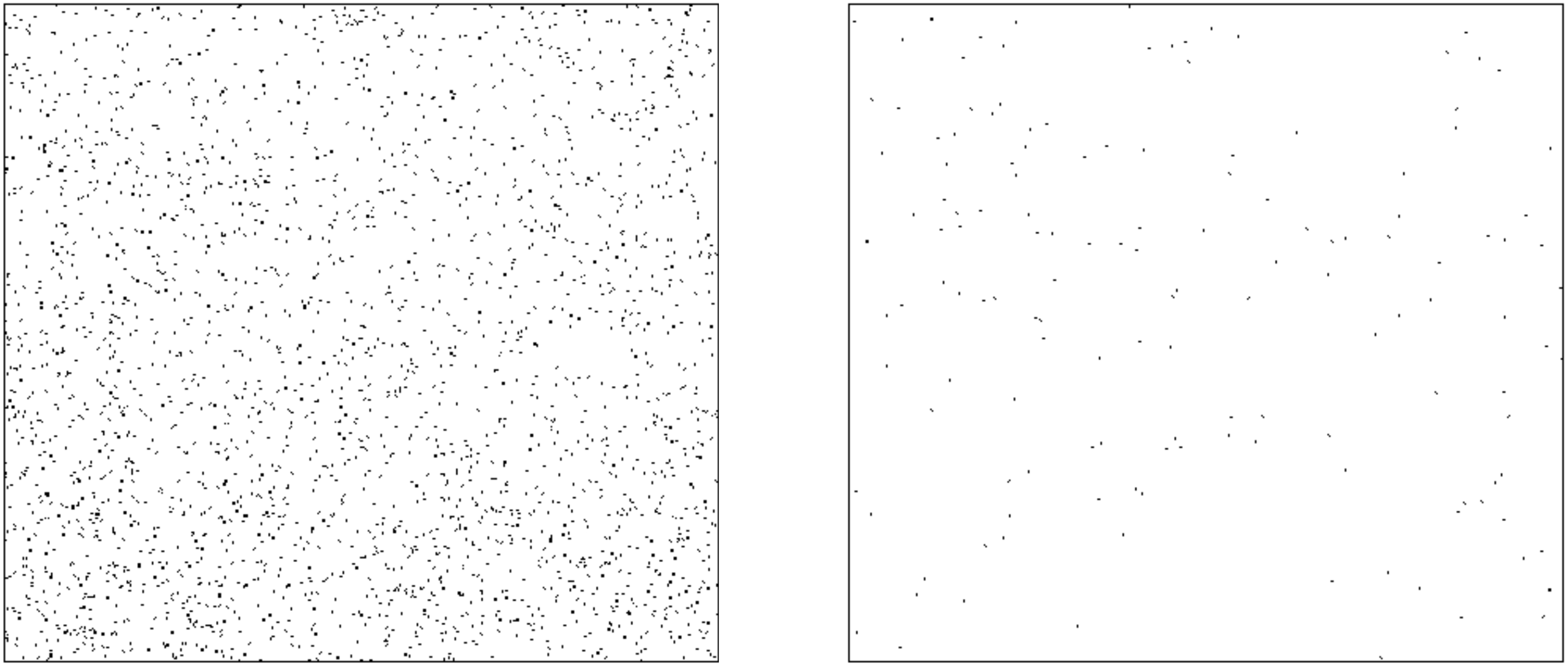}}
\caption{Phase singularities in the raw phases (experimental result). Left: Without physical averaging. Right: With physical averaging, extended light source. The charges of the phase singularities (+1 or -1) are not distinguished here.} 
\label{lPSRedDisc}
\end{figure}

Since the light source, being placed before the beam splitter, affects both arms in the same fashion, each light source point generates the same phase distribution in the detector plane. This guarantees that the systematic phase present in the data is preserved. Owing to the mutual incoherence of the light source points, the resulting  interference pattern is the sum of the intensity patterns of all the single light source points. Figure \ref{lExpRedDisc} gives the resulting raw phase so obtained. Comparing with Fig. \ref{lExpPointSource}, it can be seen that the raw phase is much less noisy, and almost appears as a result from a measurement on a smooth surface. Consequently, unwrapping with the same unwrapper used in Fig. \ref{lExpPointSource} produces only a few, insignificant errors; in fact, the most part of the raw phase has been correctly unwrapped. Figure \ref{lPSRedDisc} shows the phase singularities in the raw phase of both Fig. \ref{lExpPointSource} and Fig. \ref{lExpRedDisc}, again demonstrating the reduction in number.

\noindent

\noindent

\begin{figure}[htbp]
\centering
\fbox{\includegraphics[width=0.75\linewidth]{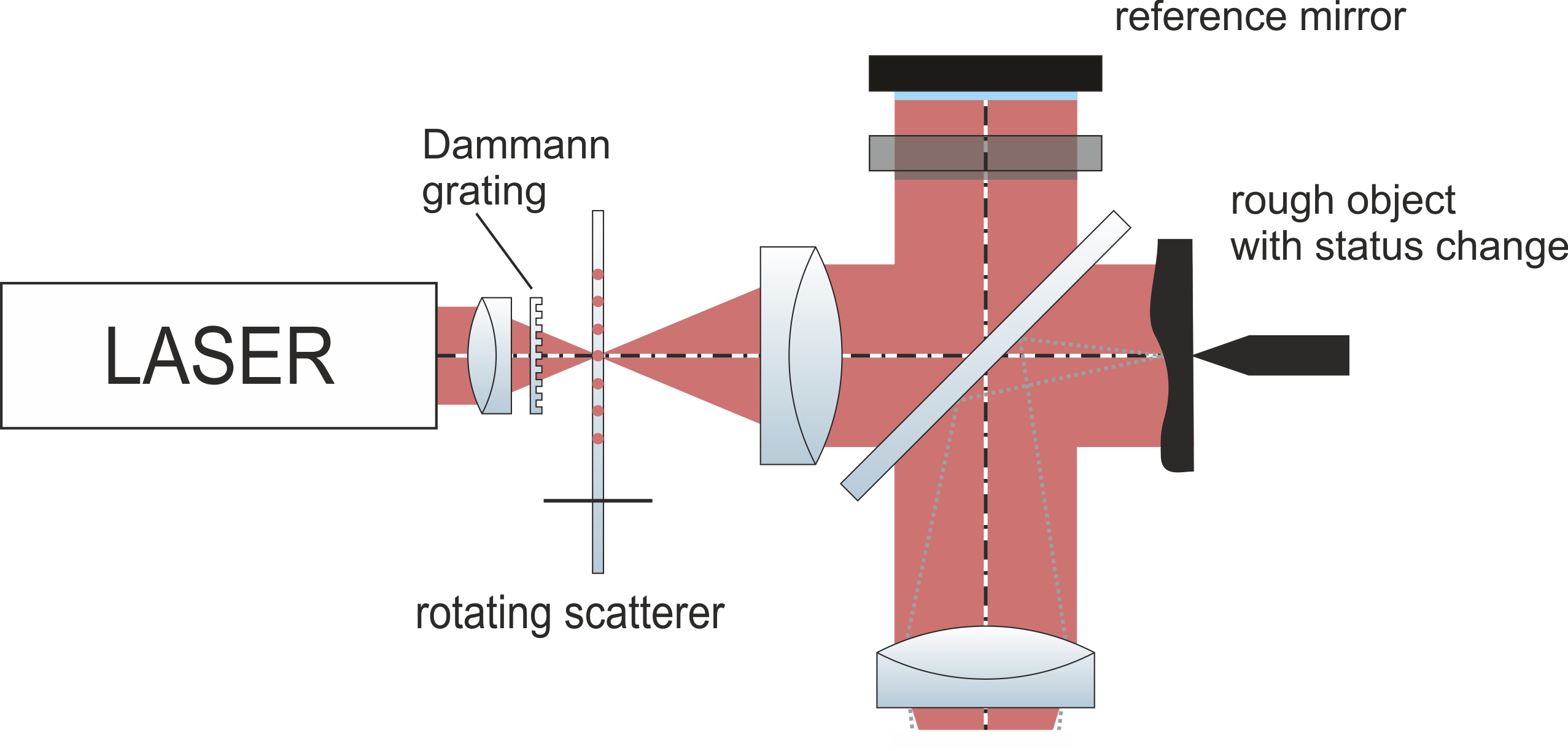}}
\caption{Alternative setup for a deformation measurement with reduced number of phase singularities. A Damman grating generates a periodic pattern on a rotating scatterer. The light source now consists of mutually incoherent, but discrete points.} 
\label{lSketchDefDamm}
\end{figure}

As was mentioned above, for an extended light source, the interferometer arms have to be adjusted to optical path difference zero, ensuring a high contrast interference pattern. This limits the allowed specimens to almost planar shape and poses a severe restriction on the setup considering flexibility and practicality. When a discrete light source is used, however, this restriction no longer holds, and averaging is still possible. Figure \ref{lSketchDefDamm} shows such a setup, where a Dammann grating \cite{Rob1992} generates a periodic spot pattern on the rotating scatterer - in our case a pattern of 5x5 points - illuminating the interferometer. The light source points are again mutually incoherent and may be spatially extended, i. e. the light need not be sharply focused on the rotating scatterer. 

\noindent

\begin{figure}[htbp]
\centering
\fbox{\includegraphics[width=0.90\linewidth]{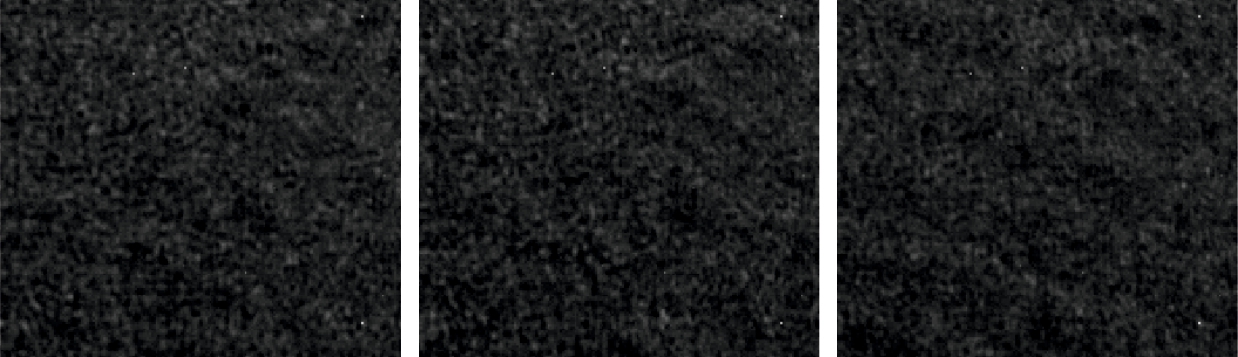}}
\caption{Speckle patterns for a discrete light source (experimental result). Left: interferometer arms balanced. Middle: imbalanced interferometer arms, OPD$\approx$2cm. Right: imbalanced interferometer arms, OPD$\approx$5cm} 
\label{lExpSpecklePattDamm}
\end{figure}

Figure \ref{lExpSpecklePattDamm} gives the speckle patterns for interferometer arms that are even stronger imbalanced than in Fig. \ref{lExpSpecklePattDisc}. Owing to the periodicity of the light source, the speckle contrast stays constant, even for an imbalance of several centimeters. 

\noindent

In contrast to the extended light source, the speckle size is now seemingly unchanged., as Fig. \ref{lExpSpecklePattDamm} illustrates. Since the following measurements show a similar reduction in the number of phase singularities, the larger speckle size for the extended light source can not be the only reason explaining this effect.

\begin{figure}[htbp]
\centering
\fbox{\includegraphics[width=0.90\linewidth]{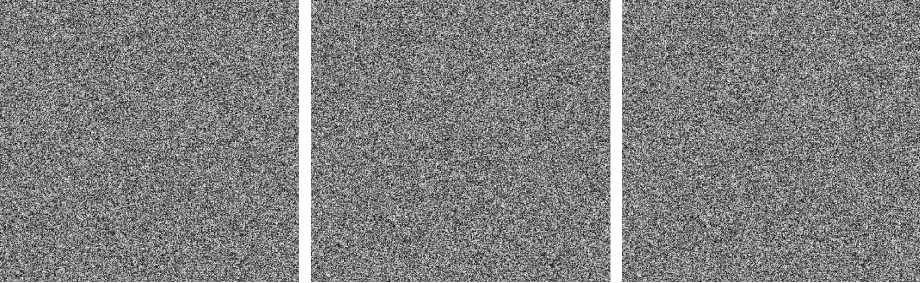}}
\caption{Speckle sizes (raw phase) for a discrete light source with varying extension (experimental result). Left: Small source. Middle: extended discrete light source. Right: Even more extended discrete light source.} 
\label{lExpSpeckleSizesDamm}
\end{figure}
\noindent

\begin{figure}[htbp]
\centering
\fbox{\includegraphics[width=0.90\linewidth]{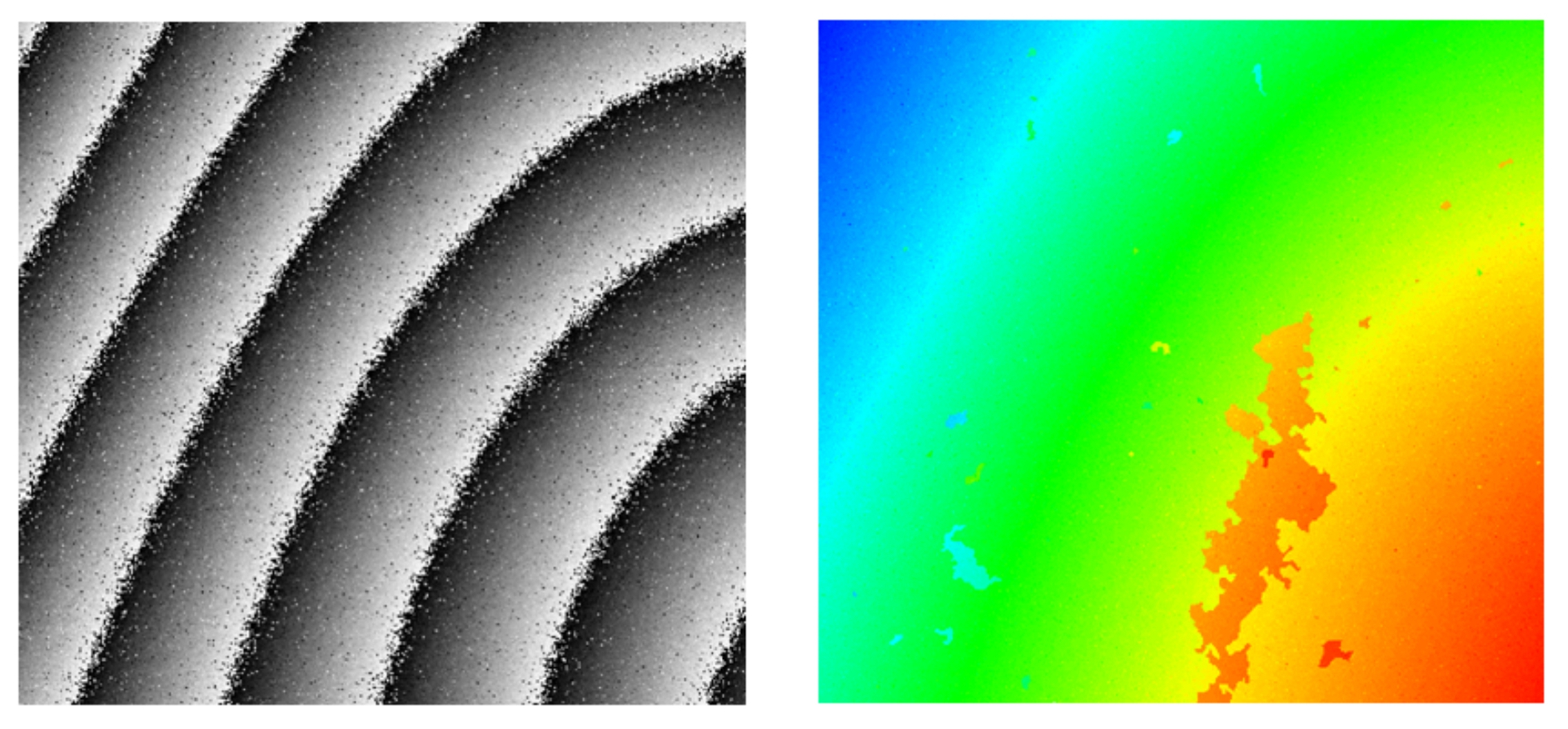}}
\caption{Results of a speckle deformation measurement, physical averaging applied (discrete light source, experimental result). Left: Raw phase (values in the interval (-$\pi$,$\pi$]). Right: Unwrapped phase (pv 7.84$\lambda$), obtained with the standard unwrapper for smooth surfaces. The number of phase singularities in the raw phase is again reduced, albeit not by the same amount as for the extended source.} 
\label{lExpRedDamm}
\end{figure}

Figure \ref{lExpRedDamm} shows the results for a deformation of the specimen similar to Fig. \ref{lExpPointSource} and Fig. \ref{lExpRedDisc}. The reference mirror has been intentionally shifted by a few centimeters to introduce a nonzero optical path difference between both arms. Again, the number of phase singularities is reduced, although not by the same amount as was the case for the extended light source. Consequently, the unwrapped phase shows more errors, although the underlying systematic phase can still be clearly seen.

\noindent

\begin{figure}[htbp]
\centering
\fbox{\includegraphics[width=0.95\linewidth]{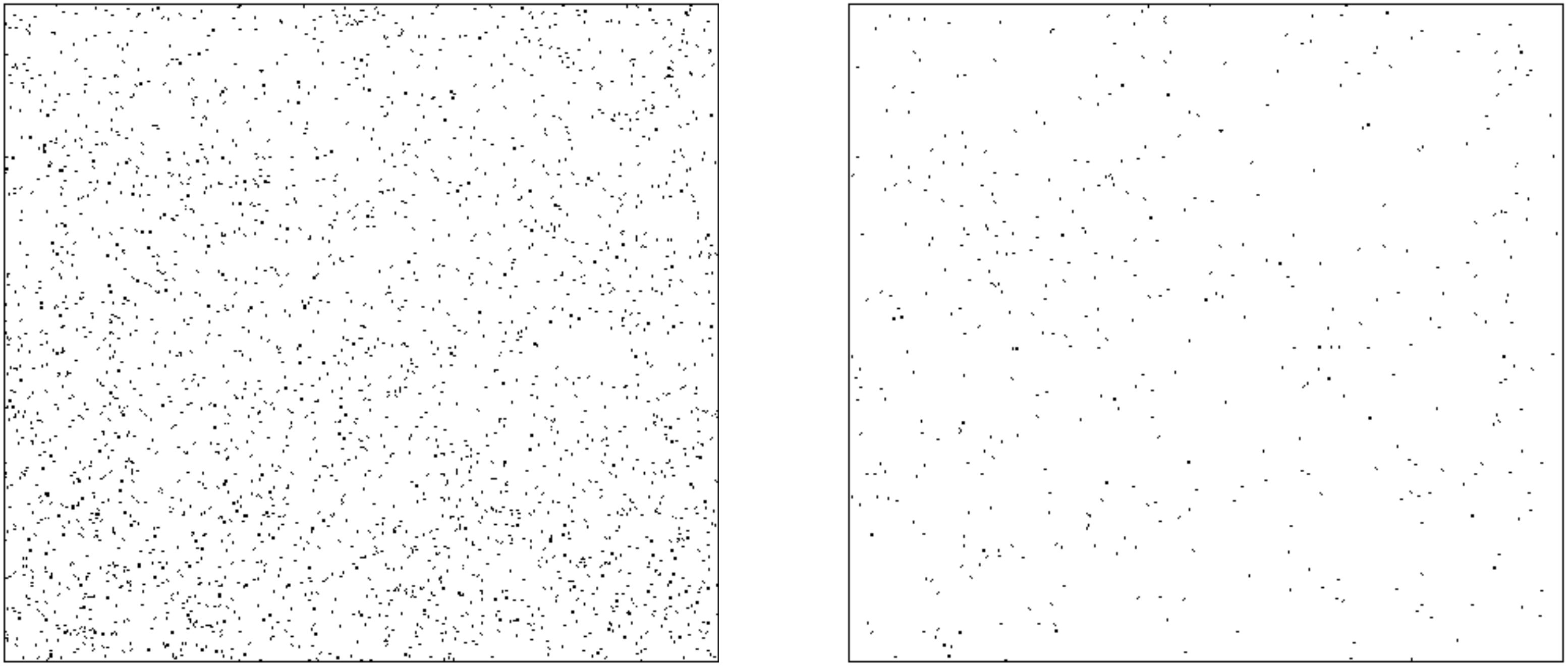}}
\caption{Phase singularities in the raw phases (experimental result). Left: Without physical averaging. Right: With physical averaging, discrete light source. The charges of the phase singularities (+1 or -1) are not distinguished here.} 
\label{lPSRedDamm}
\end{figure}

A plot of the phase singularities confirms these observations, as Fig. \ref{lPSRedDamm} demonstrates. Although the reduction in phase singularities is not as good as for the extended light source, a clear reduction in number has nevertheless been obtained.

\noindent

It should be noted that these results do not necessarily imply that an extended light source delivers better results than a discrete light source. A redesign of the Dammann grating, including an increase in light source points, might well give results with the same quality as for the extended source.

The underlying reason for the reduction in the number of phase singularities in the resulting raw phase is not fully understood at this point. Some general remarks can nevertheless be made. The reduction in the number of phase singularities due to physical averaging might take place in two different ways only: either in each of the single measurements before or after the deformation (or with wavelengths $\lambda_{\rm 1}$ and $\lambda_{\rm 2})$ separately, or while combining both measurements into the resulting raw phase. 

As for the first way, it was argued in \cite{Man2014} that such a separate reduction may be expected only in speckle fields where the phase distribution deviates from the constant one, as can be the case in speckle shearing interferometry. In particular, since the phase distribution in a Michelson setup is constant, there is no danger of destroying the systematic part of the phase as could be the case in shearing interferometry.

\begin{figure}[htbp]
\centering
\fbox{\includegraphics[width=0.70\linewidth]{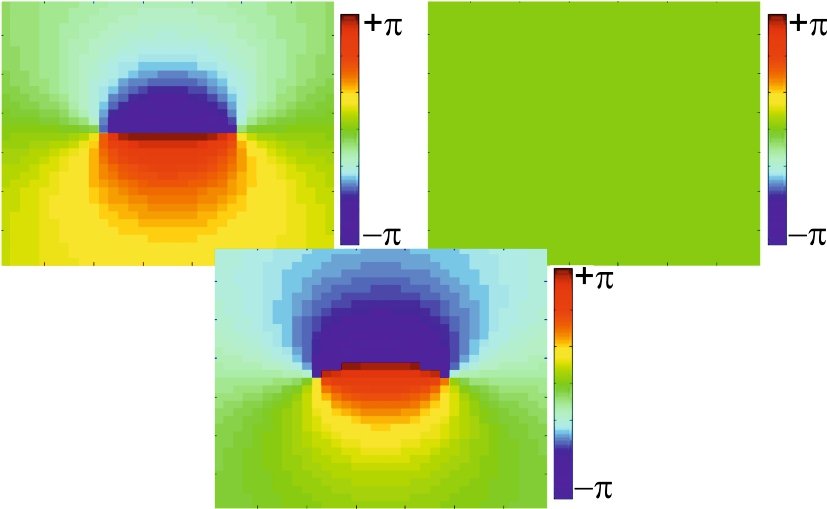}}
\caption{Combination of a phase singularity (upper left) and a region of constant phase $\pi/5$ (upper right) into the combined raw phase (bottom),simulation. The phase singularity is reproduced.} 
\label{lPSComb1}
\end{figure}

\begin{figure}[htbp]
\centering
\fbox{\includegraphics[width=0.70\linewidth]{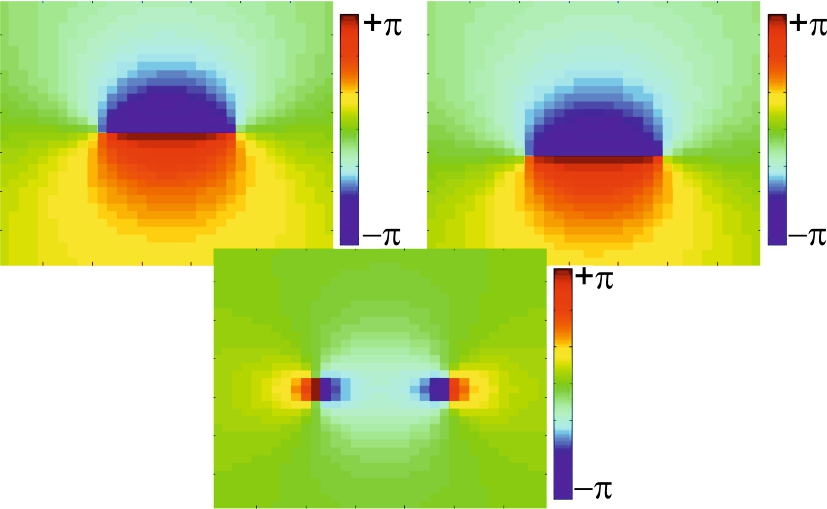}}
\caption{Combination of two phase singularities (upper left, right) into the combined raw phase (bottom), simulation. The phase singularity is mostly cancelled.} 
\label{lPSComb2}
\end{figure}

Here, it turns out that the combination of the single measurements into the combined raw phase also provides a way of reducing the number of phase singularities. With $\Phi_{\rm1}$ and $\Phi_{\rm2}$ denoting the phase maps of the two single measurements, the combined raw phase is given by

\begin{equation}	\label{eq:phi}
	\Phi={\rm atan2}\left(\frac{{\rm Im}(e^{i\Delta\Phi})} {{\rm Re}(e^{i\Delta\Phi})}\right),
\end{equation}

\noindent with $\Delta\Phi:=\Phi_{\rm2}-\Phi_{\rm1}$ denoting the difference between the single results. As Fig. \ref{lPSComb1} and Fig. \ref{lPSComb2} illustrate, combining the raw phases according to Eq. \ref{eq:phi} can reduce the number of phase singularities present in the deformation phase, provided that the phase singularity distributions are similar for both raw phases. In Fig. \ref{lPSComb1}, it is shown that a phase singularity which is combined with a region of constant phase reproduces itself. The position of the phase edge slightly changes if the constant phase is different from zero; the phase edge then follows the contour line corresponding to that value of the original phase map. 

\noindent In Fig. \ref{lPSComb2}, two phase singularities being in the same position cancel upon taking the difference. If the positions do not exactly coincide, as is actually the case in the figure, then two new phase edges appear, joining the corresponding phase singularities, but are distinctly smaller. In this way, the original singularities still (mostly) cancel. This mechanism is similar to the one shown in \cite{Man2014}, Fig. 5. 

It should be noted that, just like in shearing interferometry, the speckle fields may be replaced with vector fields associated to them. For the Michelson case, the associate vector field is given by ${\bf v}^i:=\left( Re \left\{v^i\right\}, Im\left\{v^i\right\}\right)$, $v^i:=u_O^i\cdot (u_R^{i})^*$ being the complex amplitude associated with light source point $i$, and $u_O$ and $u_R$ are the complex amplitudes of object and reference wave. In particular, the incoherent averaging of several light source points again corresponds to the addition of the associated vector fields.

The reduction mechanism mentioned above therefore only works if the averaged phase maps show an increased similarity in the position of their phase singularities, as compared to the raw phases obtained from a point source. It turns out that the physical averaging indeed increases the correlation of the phase singularity distributions in the phase maps $\Phi_{\rm1}$ and $\Phi_{\rm2}$, respectively. With a point source, the correlation coefficient, defined by

\begin{equation}	\label{eq:corr}
	c:=\frac{PS_{\rm1}PS_{\rm2}}{\sqrt{PS_{\rm1}PS_{\rm1}}\sqrt{PS_{\rm2}PS_{\rm2}}},
\end{equation}

\noindent is 0.52 for the raw phases resulting in the deformation phase shown in Fig. \ref{lExpPointSource}. Applying the extended light source, $c$ increases to a value of 0.71 for raw phases of Fig. \ref{lExpRedDisc}. The value for the discrete light source, 0.69 (Fig. \ref{lExpRedDamm}), is only slightly lower, and still significantly higher than for the point source. This higher correlation means that the chances for two singularities to cancel are increased, giving rise to the smoother and less noisy raw phases after physical averaging. In some sense, it seems that the contribution of the light source to the distribution of the phase singularities is `averaged out', leaving only the contribution of the microstructure of the surface, which is similar for small deformations. In this way, correlations are enhanced. Future research intends to make such hand waving arguments precise. 

In conclusion, it could be shown that incoherent averaging via a suitably tailored light source significantly reduces the number of phase singularities in speckle deformation measurements. It is reasonable to assume that incoherent averaging also improves other speckle techniques, like two-wavelength procedures. For the discrete light source, no balancing of the interferometer arms is necessary. The incoherent averaging is a physical process, making the procedure distinctly different from software algorithms like filtering or smoothing. 

%

\end{document}